## Can the spinning of elementary particles produce the rest energy m $c^2$? The vortex model of elementary particles

Antal Rockenbauer
*Chemical Research Center, PO Box 17, H-1525, Budapest, Hungary*

*E_mail: rocky@chemres.hu*
*Phone: 00-361-325-7900/290*

**Abstract**

The Dirac equation describes the motion of electrons in electromagnetic field, but it considers spin as intrinsic property without any real motion. We postulate spin as the intrinsic feature of vacuum, in which the incident electromagnetic radiation can create vortices with well defined spinning frequency and completely undefined axis of rotation. The vortices have finite surface determined by the spinning frequency that makes the peripheral speed equal to the velocity of light. According to the Lorentz equations of special relativity, the product of radius and mass of rotating objects is independent of the frequency of rotation and it is equal to 3/4ħ/c for all fermions. This invariance explains how the light with infinite spatial dimension and zero mass can produce finite mass and size for the particles in the pair-creation processes. The vortex model interprets the rest energy as the kinetic energy of spinning. The isotropic character of spinning can be represented by helical motion with right or left-handed chirality corresponding to the duality of matter and anti-matter. Conservation rule is proposed for the chirality explaining the electron-positron annihilation and the pair-creation processes. The vortex induced field rotation around the particle can reproduce the gravitational formula if the curvature of field is derived from the Lorentz contraction on the surface of rotating sphere. Generalized Hamiltonian is suggested for the fermions describing the spinning motion in the internal frame of particles, it reproduces the Dirac equation, when the rest energy is predominant.

## 1. Introduction

The greatest class of elementary particles, the fermions (*e.g.* electrons, protons, neutrons) can be characterized by a few intrinsic physical constants such as spin, rest mass and charge; and their motions and interactions can be described by the Dirac equation of quantum mechanics [1]. Certain physical parameters of these objects *i.e.* the radius and the moment of inertia, which are of basic importance in classical physics, are not considered to be observables in quantum mechanics, since the interactions of elementary particles with an electromagnetic field can be adequately described without defining these properties. The customarily applied point charge models avoid clarifying whether spin is related to any physical motion, and but even from the very beginning of the development of quantum mechanics there were efforts to relate certain motion to the spin [2] and time to time new proposals appeared in this subject [3-9]. These proposals, however, have never become an organic part of the main stream physics. It is partially justified, since the suggested ideas were often based on *ad hoc* and naïve concepts, no comprehensive theoretical background was given, occasionally contradicting views were expressed about quantum mechanics, the special or general theory of relativity. Without claiming to give a complete literature survey of these efforts, here we refer to a few recent papers. Levitt [3] used an argumentation based on classical physics and suggested that the spin kinetic energy contributes around 1% to the rest mass. Hestenes [4] suggested assignment of the spin kinetic energy to the zitterbewegung phenomenon of the Dirac equation, which amounts to 50% of the self-energy of electrons. Lepadatu [5] proposed the equality of the rest energy and the spinning energy of an electron rotating with the de Broglie frequency $\omega_0 = mc^2 / \hbar$ . McArthur [6] put forward the idea that the electron is a subatomic black hole, and assigned the rest energy completely to the spin. Giese [7] proposed the Basic Particle Model, when every elementary particle is built by 2 mass-less constituents, which orbit each other with the velocity of light c. This model can explain the magnetic properties of elementary particles, but it suggests spin $\hbar$ instead of $\hbar/2$. Another geometric model was proposed by Bergman [8] assuming the elementary particles are constituted by spinning charged ring, and the origin of mass and momentum were explained by electrodynamical effects, the question of spin kinetic energy, however, was not discussed. Sidharth [9] suggested a relativistic vortex model for the fermions. He based his analysis to a hydrodynamical picture, where the particlets of electron behaves like a fluid steadily circulating along a ring with radius equal to the Compton wavelength and with a velocity equal to that of light. The inertial mass is assigned to a non local term in his formalism allowing superluminal velocities, which breaks down causality. Sidharth claims this contradiction can be resolved due to the uncertainty principle within the Compton wavelength of a particle.

Though none of the papers outlined above seems to offer perfect interpretation for the concept of spin, we think there are certain ideas – in particular the hydrodynamical approach of Sidharth [9] – , what look promising to develop a less contradicting model for the electron compared to the traditional point particle approximation. In this paper we focus our attention to the major conceptual question: whether the assumption a finite size for the particles could harmonize well with the basic laws of quantum mechanics and special relativity. We retain the concept that the elementary particles cannot be divided into particlets. Though the properties of electron can be best described by quantum electrodynamics (QED), in particular the anomalous magnetic moment and the Lamb shift can be derived by very high accuracy [10], we start our analysis with the single particle approach and consider the QED contributions as a small correction in the formalism. We prefer this approach, since a conceptually well constructed single particle model could be an adequate starting point for any field theory; furthermore, this way we can avoid the conceptually awkward renormalizations applied in the theory to get rid from the imminent divergences in this formalism. We will discuss in detail if the spinning motion of particles could properly describe the basic properties of spin, if this motion could be responsible for the creation of matter and antimatter, whether the rest mass and rest energy can be connected to the spinning motion. We also investigate what could be the role of spinning motion in the processes of pair annihilation and creation. We start the analysis with an empty sphere or shell model, where correspondence is supposed between the classical rotation and the spinning motion characterized by quantum mechanical probabilities. In the subsequent part we also suggest an extension for the Dirac equation including the self coordinates of the internal motion. We visualize the fundamental properties of particles as isotropic vortices localized in the space, but we do not assume fluid particlets contrary to Sidharth [9], since according to our view the vacuum itself can be the source of mass in the spinning state. Finally we combine the vortex model and special relativity to indicate a connection with the general theory of relativity.

## 2. Discussion

### *2.1. The spinning motion of empty sphere*

#### *2.1.1. Contradictions in the point particle model*

The starting question follows: if fermions have intrinsic angular momentum, why do they not possess kinetic energy for spinning motion. The same question can be raised for photons, which have angular momentum $\hbar$, momentum $\hbar/\lambda$ and energy $\hbar\omega$, but in this case the answer is obvious: the lack of rest mass makes the definition of kinetic energy $\hbar^2/2I$ or $p^2/2m$ meaningless. Here, $I$ denotes the moment of inertia; all other symbols have the usual notation. For the non-massive photons all intrinsic physical properties are connected to the oscillating electromagnetic field. For particles with non-zero mass, we have the kinetic energy expressed by the term $(\hbar^2/2m)\Delta$ in the Schrödinger equation, but no kinetic energy is assigned to the spin in either the Pauli or the Dirac Hamiltonian [1]. The lack of kinetic energy of spinning can be rationalized if the moment of inertia vanishes, which is the case for particles with zero radius, *i.e.* when the particles are represented by point charge and point mass. The zero radius and moment of inertia, however, cannot explain how the fermions could have non-zero angular momentum, the spin. This problem can justify the effort to develop an alternative concept of quantum phenomena based on the assumption of finite particle size.

*2.1.2. Magnetic interactions in the Dirac representation*

The Dirac Hamiltonian includes the kinetic energy, the interaction with the electromagnetic field and the rest energy:

$$\mathbf{H}_D = c\vec{\boldsymbol{\alpha}}\left[\vec{p} - e\vec{A}(\vec{r})\right] + \boldsymbol{\beta} mc^2 + eV(\vec{r}) \tag{1}$$

Here, $\vec{A}$ and $V$ are the vector and the scalar potential of the electromagnetic field, respectively, while $\vec{\boldsymbol{\alpha}}$ and $\boldsymbol{\beta}$ are the four-dimensional Dirac matrices. The spin defined by the Dirac Hamiltonian has intrinsic angular momentum $\vec{S}\hbar$ with $S=1/2$, and interacts with the $\vec{B}$ external magnetic field due to its magnetic moment $\vec{\mu}_e$

$$\mathbf{H}_Z = -\vec{\mu}_e\vec{B} = -\mu_B(\vec{L} + 2\vec{S})\vec{B} \tag{2}$$

Here $\vec{L}$ denotes the angular momentum of electron orbiting around the central nucleus and $\mu_B = e\hbar/2mc$ is the Bohr magneton.

*2.1.3. Correspondence between a classical spinning object and spin*

We consider first the correspondence between the concepts of spin in classical physics and in quantum mechanics. In the classical approach, the electron can be treated as a spinning object with angular frequency $\omega$, radius $a$, charge $e$ and mass $m$. The rotation around the $z$-direction gives the angular momentum, $J_z = I\omega$ and $J_x=J_y=0$, while for the spin quantized in the $z$-direction $\langle|S_z|\rangle = \pm 1/2$ and $\langle|S_x|\rangle = \langle|S_y|\rangle = 0$. When the square of the angular momentum is calculated, the analogy fails, *e.g.* in the case of mono-axial rotation for the classical object $J^2=J_x^2+J_y^2+J_z^2=J_z^2$, while for the Dirac spin:

$$\langle|S^2|\rangle = \langle|S_x^2|\rangle + \langle|S_y^2|\rangle + \langle|S_z^2|\rangle = 3\langle|S_z^2|\rangle = S(S+1) \tag{3}$$

since

$$\langle|S_x^2|\rangle = \langle|S_y^2|\rangle = \langle|S_z^2|\rangle = 1/4 \tag{4}$$

It is noteworthy that the isotropic relation (3) is fulfilled only by $S=1/2$ spins, in other words any isotropic spinning model requires $S$=1/2 spins. Since without external magnetic field, the motion of free particles should be isotropic, we obtain a natural explanation why the elementary particles should have only ½ spin. In the classical model, where real motion is assigned to the spin, the isotropic condition can be fulfilled as a time average of the isotropic helical motion, when the average rate of rotation is the same around the axes *x, y* and *z*, and the average peripheral speed $|\vec{v}| = |\vec{\omega}|a$ is uniform over the surface of the electron. Contrary to the case of the mono-axial rotation, where only the sense of rotation can be distinguished and assigned to the magnetic quantum number ½ or – ½, different types of helices can exist, which have right-handed or left-handed chiral symmetry. In the following we will term it as self-chirality in order to distinguish this phenomenon from spin-chirality, which relates the polarization of spin to the traveling direction of electron [11].

*2.1.4. Kinetic energy of the spinning object*

A widely applied technique in quantum mechanics to treat the angular momentum components first as classical quantities, but in the final step the derived formula is rectified by substituting the values $J^2$ by $J(J+1)$. The correspondence between $J_z$ and the classical rotation around the axis $z$ by the angular velocity $\omega_{spin}$:

$$J_z = I\omega_{spin} = \pm 1/2\hbar \tag{5a}$$

Which determines $J^2$ due to the isotropic character as:

$$J^2 = 3J_z^2 = 3I^2\omega_{spin}^2 = S(S+1)\hbar^2 \tag{5b}$$

Then the spin kinetic energy $J^2/2I = 3J_z^2/2I$ can be rewritten as

$$E_{spin} = 3/2I\omega_{spin}^2 \tag{6}$$

As pointed out by Feynman [12], a simple classical model affords adequate interpretation of the gyromagnetic ratio for the orbital angular momentum in equation (2), but the doubling of the spin magnetic momentum has no classical explanation. It is noteworthy, however, that the classical model can also furnish a precise reproduction of the Fermi interaction between the magnetic moments of the nucleus and the electron [13-14]. The key point of argument is that the nucleus has finite size, and the magnetic moment obeys the dependence $1/r^3$ only outside the sphere; inside, it is constant and isotropic. This model helps avoid the singularity of the magnetic dipole energy between the electron and the nuclear moment for s-electrons, and quantitatively reproduces the Fermi term derived via the Dirac equation [15].

Naturally, in order to find correspondence between classical physics and quantum mechanics, *ad hoc* assumptions are often necessary, and thus we cannot consider any analogy as evidence for the finite size of elementary particles, but the above agreements look sufficiently encouraging. We therefore investigate what further consequences are to be expected if fermions are particles of finite size. In this case, the basic physical principles require the existence of kinetic energy for the spinning motion; since fermions possess intrinsic angular momentum, spin. For a sphere, whose mass is concentrated over the surface, the moment of inertia can be given as $2/3m\,a^2$, and the spin kinetic energy from equation (6) is

$$E_{spin} = m\,a^2\omega_{spin}^2 = m\,v^2 \tag{7}$$

Here, $v = a\omega_{spin}$ is the peripheral speed and $a$ is the radius of electron.

*2.1.5. Equivalence of spin kinetic energy and rest energy*

Equation (7) can be used as a starting point for various physical concepts. If we assume zero peripheral speed and, consequently, zero angular speed, the spinning motion will have zero kinetic energy. Actually this assumption corresponds to the point charge model, but in this case the origin of spin cannot be explained: how the elementary particle could have intrinsic angular momentum, when the moment of inertia is zero and no oscillating electromagnetic field is present? As a second possibility, we can assume the equality of spinning frequency and the $\omega_0 = m\,c^2/\hbar$ de Broglie frequency. In this case equations (5) and (6) will give $E_{spin} = 3/4m\,c^2$ that is 75% of the rest energy can be accounted for as the spinning energy. This assumption, however, still keeps obscure the origin of the remaining 25% of the rest energy. We prefer for this reason a third concept, when **the rest energy can be produced in full by the spinning motion of elementary particles if the peripheral speed is equal to the velocity of light**:

$$E_{spin} = m\,c^2 \tag{8}$$

In this case, equations (5) and (6) can also give the angular velocity $\omega_{spin}$ and the radius of the electron:

$$\omega_{spin} = 4/3m\,c^2/\hbar = 4/3\omega_0 \tag{9}$$

and

$$a = c/\omega_{spin} = 3/4\hbar/m\,c = 3/4\lambda_c \tag{10}$$

where $\lambda_c$ is the Compton wavelength. This result is analogue with the vortex model of Sidharth [9], only the numerical factor ¾ is different.

According to equation (10), the radius of electron is determined by the special relativity, which gives upper limit for the peripheral speed. This concept offers obvious explanation why the electron is not exploded due to the electrostatic repulsion of charge: the escape of the parts of electron would require a speed exceeding the speed limit. We consider this fact as evidence for the indivisibility of electron, it cannot be decomposed into smaller particlets. A peculiar property of equation (10) is that the frequency of spinning, $4/3\omega_0$ is larger than the de Broglie frequency characterizing the wave nature of particles, but smaller than $2\omega_0$, the level necessary for the colliding photons to create a particle-antiparticle pair.

*2.1.6. Explanation for the anomalous magnetic moment of spin*

The circular currents produced by the electron spinning with the angular speed $\omega_{spin}$ produces also a magnetic moment. Straightforward calculations show, that this moment is just half of the value derived from the Dirac equation: $\vec{\mu}_{el} = \mu_B \vec{S}$ , but the magnetic interaction is extended by an additional 'mass' effect. Since the Larmor precession will be superimposed on the spinning rotation, the rest mass and rest energy will be also changed. The relation between the magnetic dipole interaction and the change of spinning kinetic energy can be given analogously with equation (8):

$$-\mu_B \vec{S}\,\vec{B} = \Delta m\, c^2 \qquad (11)$$

The sum of these direct (charge) and indirect (mass) effects doubles the strength of magnetic interaction of spin. Since the classical model of rotating sphere does not take into account the double impact of magnetic field, the complete formula obtained from the second order perturbation treatment of Dirac equation [15] cannot be reconstructed without supposing that the rest energy is produced by the kinetic energy of spinning particles. A small further contribution to the magnetic moment ($\alpha\mu_B/\pi$) is derived by the QED describing the phenomenon of vacuum polarization [10]. The combination of QED with the vortex model of single particles may offer a deeper understanding of the very nature of elementary particles, but it requires further clarification of the complex scheme of interactions between particles and the electromagnetic field, which question is outside of the scope of present paper. Here we only remark the extension of equality in (11) by the contribution of vacuum polarization can reconstruct precisely the experimentally observed giromagnetic ratio.

*2.1.7. Compton radius of electron in the Dirac formalism*

The perturbation solution of the Dirac equation supports also the assumption for the finite size of electron radius given by equation (10) stating it is nearly equal to the Compton wavelength $\lambda_c$. **When the relativistic and non-relativistic terms are separated, one defines the electron not as a point charge, but as a distribution of charge and currents extending over a domain of linear dimension with** $\lambda_c = \hbar / mc$ [16].

*2.1.8. Rest energy and spin kinetic energy*

The rest energy $mc^2$ can be expressed by $\omega_{spin}$, or the radius of electron, or by the moment of inertia:

$$E_{spin} = 3/4\hbar\omega_{spin} = 3/4\hbar c/a = S(S+1)\hbar^2/2I \qquad (12)$$

The *S(S+1)* dependence of the spinning kinetic energy shows the analogy with the rotation energy of molecules in the rotational spectroscopy. In equation (12) $\hbar c$ can be substituted by the square of elementary charge $e^2 = \alpha\hbar c$, where $\alpha = 1/137$ is the Sommerfeld's fine structure constant. This expression can be extended to the quarks by introducing the *n* charge quantum number: $e^2 = \alpha n^2 \hbar c$, where $n$=-1/3 for the down-quark, $n$=2/3 for the up-quark, $n=-1$ for electron, $n=1$ for positron, $n=0$ for the neutrinos, respectively [17-18]. Then the rest energy of massive particles can be rewritten in the analogy of the electrostatic interaction:

$$E_{spin} = \frac{3}{4\alpha n^2} e^2 / a \qquad (13)$$

*2.1.9. Invariants in the special theory of relativity and the vortex model*

Though the special theory of relativity deals with only inertial systems, we can extend the formalism to rotating system if the peripheral speed is constant and we decompose the motion into infinitesimal parts. By introducing polar coordinates, the contraction of perimeter can be given by the Lorentz formula. Denote by $a_s$ and $m_s$ the radius and the mass in the self-system of spinning particles, while *a* and *m* are the same quantities 'observed' in the external system of inertia:

$$m_s = m(1 - v^2/c^2)^{1/2} \quad \text{and} \quad a_s = a(1 - v^2/c^2)^{-1/2} \qquad (14)$$

There is a peculiar consequence of equation (14), namely the peripheral radius $a = perimeter/2\pi$ differs from the central radius $R$ of the sphere, since no motion exists in radial direction, and consequently, no Lorentz contraction can occur. In other words the concept of rotating sphere involves *a priori* non-Euclidean geometry. Though the Lorentz transformation was originally introduced for macroscopic objects, the successes of Dirac equation clearly prove the validity of Lorentz formulas for quantum mechanical objects. In this latter case, however, we have to apply quantum mechanical rules to describe the isotropic rotation: the axis of rotation points into all direction with the same probability. Thus the expectation value of radius $a$ should be the same in all directions, and the peripheral radius $a$ can be defined by the relation: $surface = 4a^2\pi$.
In the limiting case $v \to c$ equation (14) gives asymptotically zero self-mass and infinitely large self-radius. Consequently, when the self-system is considered, **the elementary particles behave like empty space (vacuum), and we can postulate the elementary particles as vortices defined as spinning confinements of the space.** This concept offers new interpretation for the phenomena of electron-positron annihilation and pair creation. In the annihilation process, the particles with zero self-mass and infinite self-dimension produce electromagnetic radiation with exactly the same physical characteristics with the exception of spin (See Table 1).

*2.1.10. Chirality of spinning and the matter-antimatter duality*

In the pair-creation process, the electromagnetic radiation is transformed into the particle state, where the spinning of a particle with the velocity of light creates the finite mass and restricts the dimension to a finite size. Since the isotropic spinning motion can be represented by helices either right- or left-handed, we can assign the matter-antimatter duality to the two senses of chirality. By defining the self-chirality as additive quantity with $K_C$=1 for the matter, $K_C$= -1 for the anti-matter and $K_C$=0 for the radiation, respectively, we can introduce conservation law for the self-chirality that can govern the processes between matter and anti-matter. If two electrons or two positrons collide, the $K_C$ values are added and no mass annihilation can occur. On the other hand, annihilation can take place when an electron and a positron collide, since in this case, the opposite sign of self-chirality results in zero $K_C$ and, consequently, the mass can be transformed to radiation. The decay process of neutron is also in agreement with the conservation of self-chirality, since the neutron decays into two particles (an electron and a proton) and one antiparticle (antineutrino).

Since the spinning motion does not shrink the central radius of vortices, the particles can interact in any distance being their central radius infinitely large and it can give rise to uniform strength of interaction. The electric charge can be interpreted as the source of repulsion between vortices of the same chirality, or attraction if the chirality is opposite and the spinning frequency is the same, like in the cases of electron and positron, or proton and antiproton, respectively. The latter example, however, represents composite particles. The strength of repulsion/attraction is the same for elementary vortices even for the case when the spinning frequency is different (electron, muon and tau), the only exceptions are represented by the quarks. Maybe the confinement of quarks is related to the fractional charge: they can exist only inside of composites having integer charge. In other words the quarks can be defined as non-observable particles. Since the chirality defines simultaneously the matter-antimatter duality and the sign of electric charge, the charge conjugation is identical with the $K_C$ parity of internal space converting the right- and the left handed chiralities. The Pauli exclusion principle can be expressed also in terms of self-chirality: the particles with identical handedness could never be in the same quantum mechanical state, since in this case the repulsion would be infinitely large. *E. g.* when the Coulomb interaction is computed between two electrons differing only in the magnetic quantum number, the position of particles is described by independent probabilities. The concept of independent spatial motion of vortices would be invalidated if all quantum numbers were the same for two electrons. Alternatively, the Pauli exclusion principle can be expressed also by the anti-symmetry of wave function when two electrons are interchanged. For positronium the $K_C$ charge conjugation is equivalent with the interchange of electron and positron. In this case the wave function should be also anti-symmetric for $K_C$ to keep independent probability distribution for the two vortices, since the identical spatial position of electron and positron would create infinitely strong attraction resulting in prompt annihilation of the vortices.

*2.1.11. Invariants of fermions*

Table 1 compares the intrinsic constants of photon and selected fermions. As the table shows the particles in their self system have the properties of a real vacuum without mass, charge, spin and spatial confinement, but the spinning with adequately large frequency can fill the vacuum with the characteristic features of particles. The primary feature of vortices is the spinning frequency, which can determine all intrinsic parameters of the particles. If it is zero, neutrino or antineutrino is created with zero mass (see the Standard Model [17-18] ) and the peripheral radius of neutrino is infinite. By increasing spinning frequency the particles with increasing masses and decreasing peripheral radius can be formed (electrons, muons, baryons).

For any fermions (like the nucleons) the identity of spin kinetic energy and rest energy can be valid, since all have the same spin momentum and only the rest mass and the charge differ. The relativistic invariant of spinning particles

$$m_s a_s = m\,a = 3/4\hbar/c \tag{15}$$

can be regarded as a universal constant of all fermions, which expresses the fact that all fermions have the same spin and the peripheral speed is equal to *c*. It means the velocity of light is not only a universal constant of the special relativity, but also a fundamental invariant of the elementary particles. The invariance of the product of particle mass and radius represents the potentiality of electromagnetic radiation with zero mass and infinite spatial dimension to be converted into particles with non-zero mass and confined size provided by the incident photon has adequate energy to form the spinning motion, where the condition $m = 3/4\hbar\omega_{spin}/c^2$ can be fulfilled and $a = c/\omega_{spin}$. In other words, the spinning motion can convert each other the two basic entities of material. If the incident photon has the energy $\hbar\omega \geq 3/2\hbar\omega_{spin}$, a pair of the spinning particles with opposite chirality can be formed. The driving force of this process is the conversion of $\hbar$ angular momentum of the circularly polarized electromagnetic radiation into the spin angular momentum $1/2\hbar$ of the created fermion pair.

According to the isotropic spinning model, the moment of inertia $I = (3/8m)(\hbar/c)^2$ is inversely proportional to the mass, *i.e.* in the self-system *I* is infinitely large. The product of the moment of inertia and the mass can be also expressed by the universal constant of elementary particles:

$$I_s m_s = I\,m = 3/8(\hbar/c)^2 = 2/3(m\,a)^2 \tag{16}$$

*2.1.12. Scattering data and the radius of particles*

The question can be raised if the radius of particles in Table 1 can be related to the cross section data of scattering experiments. Thomson scattering is the scattering of electromagnetic radiation by a charged particle. In this case primarily the electric component of incident wave accelerates the particle, which makes the cross section proportional to the square of classical electron radius $e^2/mc^2$. This radius, however, cannot be considered as the 'observed' radius of electron, since it is two orders of magnitude smaller than the de Broglie wave length, which is - according to the uncertainty principle - the lower limit of size measurements (Note, any radiation of higher energy corresponding to this wave length would create new particles). The electron scattering gives, however, information to the size of protons, since the scattering cross section is smaller for the spread-out charge and current cloud compared to the point particle [19]. By using exponential charge and current distribution, the root-mean-square radius was found around 5 times larger than the value of peripheral radius given in Table 1. In the case of shell model this ratio is reduced to a factor of 3, which is still too large to explain the deviation by experimental uncertainties. The reason of deviation lies in the composite character of proton since the orbiting motion of quarks can significantly increase the overall radius of charge distribution. Consequently no direct experimental data can be given to the radius of any really elementary particle, and thus we consider the peripheral radius only as an auxiliary parameter that can give a unique relation between the angular momentum, the rest energy and the magnetic moment of fermions.

*2.1.13. Connection between the special and general theory of relativity*

It is tempting to assume that the most general interaction, the gravitation is also linked to the spinning motion: its strength depends on the curvature of space, which in turn is proportional to the spinning frequency $\omega_{spin}$. This relation follows from the vortex model since its peripheral rotation will separate the space into two parts, a rapidly spinning part with the size $a=c/\omega_{spin}$ (we call it as "particle") and the surrounding of this particle, where slow 'gravitational' rotation with the frequency $\omega_{gr}$ is induced due to the discontinuity of space created by the vortex. Assume the gravitational frequency decreases with the *R* distance from the vortex according to the law:

$$R^3\omega_{gr}^2 = 2f\,m \tag{17}$$

where *f* is the gravitational constant. The Lorentz contraction will reduce the self-perimeter of sphere $2R\pi$ to the contracted value $2r\pi$:

$$2r\pi = 2R\pi\sqrt{1-\frac{r^2\omega_{gr}^2}{c^2}} \tag{18}$$

In the case of small contraction, the curvature of non-Euclidean space can be defined by $c^2(1-r/R)$, which can be expressed by $\omega_{gr}$ if we keep only the first order term in the Taylor expansion of equation (18):

$$curvature = c^2(1 - r/R) = R^2\omega_{gr}^2/2 \qquad (19)$$

This definition gives zero curvature if $R$=$r$. Comparison of equations (17) and (19) gives the curvature $f\,m/R$ at the distance $R$ from the $m$ mass, that is the gravitational potential corresponds to the Newton's law between the masses $m$ and $M$

$$V_{gr} = -f\frac{mM}{R} \qquad (20)$$

When the distance $R$ is equal to the radius $a$ given by Table 1, the gravitation force is too small to fulfill the assumption of McArthur [6], who considered the elementary particles as black holes.

*2.2. The quantum mechanics of spinning motion*

*2.2.1. The generalized fermion equation*

Obviously the self motion of particles – like the orbital motions of electrons – has non-classical nature and it should be considered as a stationary state of spinning particles. While Dirac equation describes the particle motion only in the space of position vector $\vec{r}$ , we have to introduce also the internal coordinates of particles describing the space, where the self-rotation takes place. We can define the internal-radius $\hat{a}$ of the particle as an operator and introduce three quantum numbers for characterizing the type of self-rotation. The actual particle size '$a$' is considered as the eigenvalue of the size operator $\hat{a}$ , the quantum number $M$=1/2 or -1/2 will distinguish the two spin states, the $\beta$=1 or -1 the positive and negative energy states, and finally the $K_C$=1 or -1 chirality index is assigned to the particles and anti-particles, respectively. This notation allows extending the Dirac equation for all fermions. The particle charge could have the value $n\,e$, where the $n$=1, -1, 0, 2/3, 1/3. In order to consider both particles and antiparticles as two distinct states of self-rotation, we extend the 4 dimensional spinor field of Dirac to the 8 dimensional spinor representation ρ:

$$\rho_{\alpha,i} = \begin{pmatrix}\alpha_i & 0\\ 0 & \alpha_i\end{pmatrix} \text{ and } \rho_\beta = \begin{pmatrix}\beta & 0\\ 0 & \beta\end{pmatrix} \qquad (21)$$

$$\rho^*_{\alpha,i} = \begin{pmatrix}\alpha_i & 0\\ 0 & -\alpha_i\end{pmatrix} \text{ and } \rho^*_\beta = \begin{pmatrix}\beta & 0\\ 0 & -\beta\end{pmatrix} \qquad (22)$$

where i=x,y,z.

By applying equation (15) describing the relativistic invariance $m\,a=¾\hbar/c$, we can derive the kinetic energy of particle with spin $J$ and moment of inertia $I=2/3m\,a^2$:

$$\mathrm{H}_{spin} = J^2/2I = \frac{3}{4}\hbar^2/(\frac{4}{3}ma^2) = \frac{3}{4}\hbar c/\hat{a} \qquad (23)$$

By replacing the rest energy in the Dirac equation by equation (23) and extending the energy expression by the predominant (one-loop) QED contribution $\alpha/\pi$, we obtain the generalized equation of fermions:

$$\mathrm{H}_F = c\left[\vec{\rho}_\alpha\vec{p} - \vec{\rho}^*_\alpha n\,e\,\vec{A}(\vec{r})\right] + \frac{3\rho_\beta\hbar c}{4\hat{a}}(1+\alpha/\pi) + \rho^*_\beta n\,e\,V(\vec{r}) \qquad (24)$$

The coordinates and quantum numbers of the extended wave function is $\Psi(\vec{r}, a, M, K_C, \beta)$, where the variables of the position and self-rotation can be separated if the spin kinetic energy (the rest energy) is large compared to the other terms. In this case we can use the first order perturbation approach and diagonalize the spin kinetic energy. The eigenfunction of the spin kinetic operator is a delta function $\psi(a') = \delta(a'-a)$, where '$a$' is the particle radius, and the eigenvalue of kinetic energy $¾\hbar\, c/a=m\,c^2$ is equal to the rest energy according to equation (15). The angular frequency $\omega$ in the stationary wave function $exp(i\omega t)\ \psi(a)$ is equal to the de Broglie frequency $m\,c^2/\hbar=¾c/a$, which is smaller than the spinning frequency defined as the ratio of the $z$-component of the angular momentum (spin) and the moment of inertia: $\omega_{spin}=J_z/I=c/a$. The expectation value of the other terms trivially gives the usual Dirac equation, since these terms are independent of the variable $a$, and consequently the Dirac equation can be considered as the first order approach of the generalized Hamiltonian of fermions. Since Hamiltonian (24) includes also the QED contribution, the precise values of Lamb shift and the gyromagnetic ratio are reconstructed. Furthermore, the operator $\hat{a}$ has exact eigenvalue $a$, thus the momentum of spinning motion should be undefined according to the uncertainty principle. It is indeed the case, since only the absolute value of peripheral speed is known (the velocity of light $c$), while its direction is undefined.

The inclusion of internal coordinates offers a straightforward explanation for the parity violation in the beta decay process of neutrons. The asymmetry takes place since the definition of parity is not complete: only the external coordinates of position are reflected. If besides this reflection $K_P$, charge conjugation $K_C$ is also applied, the beta decay is no more asymmetric. The double reflection $K_PK_C$ can be substituted in our model by the simultaneous reflection of external and internal coordinates changing also the sense of self-chirality, which in turn interchanges the particles and antiparticles.

The spin kinetic energy in the generalized Hamiltonian (24) varies continuously as a function of mass, thus our model - like the Standard Model of elementary particles – cannot give answer for the fundamental question: why the masses of free particles can have only discrete values. Further shortcoming of the Hamiltonian (24) is the arbitrary introduction of charge quantum number $n$: in the consequent operator formalism one should derive all quantum numbers by diagonalization of the complete Hamiltonian. Resolution of the above inconsistencies may be expected by introducing second, or higher order terms in the Hamiltonian [20] or by the development of unified field theory for the elementary particles including both the electro-weak and strong forces.

In the high energy domain, when the motional and rest energies are comparable, one expects strong coupling between the self-rotation and the orbital motion. In this case, one can obtain deviations from the usual second order perturbation terms of the Dirac Hamiltonian and these deviations could give a chance for verifying experimentally the suggested Hamiltonian (24) of fermions.

### *2.2.1. Fundamental forces and the frequency of spinning*

In our model the discrete values of particle mass are related to the discrete spinning frequencies creating finite discontinuities in the space. We presume the four fundamental forces affecting the motion of particles are generated at various level of spinning frequency and we distinguish altogether three types of particles. Below the first critical frequency level neither electromagnetic, nor strong forces can affect the motion of particles: these particles are the neutrinos, which interact only with the weak forces. Above the first critical frequency level, the particles (electrons, positrons, muons) can possess the electromagnetic interaction, but the strong interactions are still not available. When the second critical level is achieved by the spinning frequency, the particles (quarks, baryons) can also interact by the strong forces. The large spinning frequency is necessary, but not always sufficient for the creation of color charge mediating the strong interaction, as it is exemplified by the heavy tau particles in the third generation of electron family. Our classification differs from the one applied in the Standard Model, since we categorize the particles according to the number of activated forces, and as a consequence the leptons are divided into two classes: the “weak” (neutrino) and the “electro-weak” (electron) particles. All the three classes of particles have three generations and the higher the index of generation the larger is the spinning frequency.

## 3. Conclusions

We made an attempt to give new interpretation for the intrinsic properties of fermions defined by the Dirac equations. Instead of the generally accepted standpoint postulating the elementary particles as point-like objects, we suggest an alternative concept, where the elementary particles are vortices that are considered as local rotational states of the vacuum extending over finite surface and having isotropic spinning. The mass and charge are global and indivisible properties of vortices. This concept renders the rest energy to the kinetic energy of spinning, which can explain why the spin angular momentum compared to the orbital momentum gives twice as large contribution to the Zeeman energy: the magnetic field induced Larmor precession modulates the spinning frequency of particles, which in turn alters the kinetic energy of spinning electron, and the two effects double the overall impact of magnetic field.

The vortex model postulates the peripheral speed of spinning as equal to the velocity of light, since this speed is allowed only for massless objects, the particles should have zero mass and infinite size in their self system. These quantities have asymptotic values giving finite product of the mass and radius of particles, and the product is invariant for the rotating shell according to the Lorentz rules of relativity.

When the electromagnetic irradiation creates pairs of particles or *vice versa,* the annihilation process produces electromagnetic wave, the key role is played by the intrinsic angular momentum, the spin: As concerning to the electromagnetic wave, the oscillating electric and magnetic field carrying the spin $\hbar$ rotate in the infinite space, while in the case of particles, the space executes a local rotation and this rotation is observed in the ‘external’ frame as a particle with finite mass, finite size and angular momentum $1/2\ \hbar$. In this model the particles can be visualized by an isotropic vortex extended in two dimensions (the surface is finite, while the central radius is infinite), contrary to the string concept defining the particles as one-dimensional extended objects. The characterization of spinning as helical motion offers a chance to interpret the duality of material and anti-material as the chirality of spinning. This type of chirality satisfies a conservation rule explaining the formation of antineutrino in the β-decay of neutron. The handedness of chirality is inversed by the reflection of internal particle coordinates, thus the combination of external parity $K_P$ and charge conjugation $K_C$ can be interpreted as the full parity operation including reflection both in the external and internal coordinates. Thus in the β-decay process of neutron, the parity violation is caused by the incomplete definition of classical parity.

In the case of slow or asymptotically zero frequency of the vacuum rotation, the vortices represent the neutrinos with very small, or zero mass. This particle has no charge: it means a critical spinning rate is necessary for the particles to possess the electromagnetic interactions. If the first critical spinning rate is achieved, the particles of the electron family are created. There is a second critical spinning frequency, when the heavy and composite particles can be produced, then the higher level of frequency creates a complex pattern of vortices possessing also strong interaction. These particles are the baryons or hadrons including the quarks, but in the latter case the masses are unknown due to the phenomenon of confinement. The existence of three generations of particles indicates that the enhancement of spinning frequency does not necessarily yield a new class of particles, in other words, the various classes of particles have always two excited states.

The combination of vortex model and special theory of relativity can reproduce the gravitational formula, if we assume induced rotation outside of the vortex, where the frequency of external rotation decreases by the distance and proportional to the square root of the mass. Then according to the Lorentz contraction, the curvature of space will be inversely proportional to the distance in an agreement to the gravitational potential between masses.

This model offers a new answer for the question why the two faces of mass: the inertia and the gravitation are identical. In the general theory of relativity this identity follows from the proportionality between the curvature of space and the mass [21], while in our model the identity is caused by the common origin of these entities: both the mass and the curvature of space are determined by the spinning frequency.

According to the classical quantum mechanics, the radius and the moment of inertia of elementary particles are not physical observables. This point of view reflects the fact, that the measurement of electron radius would need so high radiation energy, where $\omega$ exceeds the de Broglie frequency, and under these conditions the particles cannot preserve any more their existence. The quantum mechanical description of elementary particles can be extended if we introduce internal coordinates for the spinning beside the usual external coordinates. By making use the relativistic invariant for the product of mass and radius for a spinning shell, the generalized Hamiltonian can be suggested for the fermions, where the eigenvalue of internal radius can characterize the various particles. This Hamiltonian reproduces the Dirac equation, if all energy terms are small compared to the rest energy. In the energy domain, where the rest energy is comparable to the other terms, there is a chance to devise experiments to check the validity of our model based on the self-rotation of particles.

Throughout in the paper, we discussed only the fermions, but the vortex model can be extended to other particles e.g. for bosons that according to the Standard Model can be built up from the combination of fermions. Since the Dirac Hamiltonian is the quantum mechanical adaptation of the relativistic equation of motion, we may also assume that the rest mass and energy of any composite object are produced by the spinning of the elementary particles. The mass deficit of composite objects can be interpreted as the slowing of spinning rotation when the particles are bond together. In the fusion or fission processes of atomic nuclei, the vast energy of escaping irradiation is supplied by the partial loss of the spinning kinetic energy of nucleons.

## 4. References

[1] P.A.M. Dirac, Principles of Quantum Mechanics. Clarendon, 1962.
[2] G.E. Uhlenbeck and S. Goudsmit, Naturwissenschaften, **47** (1925) 953.
[3] L.S. Levitt., Lettere Al Nuovo Cimento, Series 2., **34 (**1982) 333.
[4] D. Hestenes, Found. Physics, **15** (1983) 63.
[5] C. Lepadatu, Journal of Theoretics, **V-1** (2003) 1.
[6] W. McArthur, The General Science Journal, (2005) 21.
[7] A. Giese, The Seeming Mistery of the Electron, ag-physics, 2008.
[8] D.L. Bergman, Hadronic Press, Supp., Proc. Phys. Sci., Cologne, (1997), Common Sense Science 1
[9] B.G. Sidharth, in quant-ph/9808020 (1998).
[10] G. Gabrielse, D. Hanneke, T. Kinoshita, M. Nio, B. Odom, Phys. Rev. Lett., **97** (2006) 030802.
[11] G.L. Kane, Modern Elementary Particle Physics, Perseus Books, 1987.
[12] R.P. Feynman, R.B. Leighton, M. Sands, The Feynman Lectures on Physics, Vol. 2. Addison-Wesley Publishing, Massachusetts; 1964.
[13] C.P. Slichter, Principles of Magnetic Resonance, Harper and Row, New York, 1963.
[14] R.A. Ferrel, Electron Spin Resonance, Theory and Application, John Wiley et Sons, New York, 1973.
[15] A. Messiah, Quantum Mechanics, Vol. 2, North-Holland, Amsterdam, 1962, p. 948.
[16] L.L. Foldy, S.A.Wouthuysen, Phys Rev., **78** (1950) 29.
[17] D.J. Griffith, Introduction to Elementary Particles, John Wiley et Sons, New York, 1987.
[18] L.K. Gordon, Modern Elementary Particle Physics, Perseus Books, 1987.
[19] R. Hofstadter, Rev. Modern Phys., **28** (1956) 214.
[20] A.O. Barut, Phys. Lett., **73B** (1978) 310.
[21] I.R. Kenyon, General Relativity, Oxford University Press, 1990.

Table 1. Spinning frequency and intrinsic parameters for particles.

| Particle | $\omega_{spin}$ (Hz) | Self-chirality | Spin | m (kg) | a (m) | Charge | Peripheral speed |
|---|---|---|---|---|---|---|---|
| Photon | 0 | 0 | 1 | 0 | ∞ | 0 | 0 |
| Particles[a] | 0 | 0 | ½ | 0 | ∞ | 0 | 0 |
| Neutrino | 0 | 1 | ½ | 0 | ∞ | 0 | 0 |
| Antineutrino | 0 | -1 | ½ | 0 | ∞ | 0 | 0 |
| Electron | $1.036 \times 10^{21}$ | 1 | ½ | $9.109 \times 10^{-31}$ | $2.896 \times 10^{-13}$ | -e | c |
| Positron | $1.036 \times 10^{21}$ | -1 | ½ | $9.109 \times 10^{-31}$ | $2.896 \times 10^{-13}$ | e | c |
| Proton[b] | $1.902 \times 10^{24}$ | 1 | ½ | $1.672 \times 10^{-27}$ | $1.578 \times 10^{-16}$ | e | c |
| Antiproton[b] | $1.902 \times 10^{24}$ | -1 | ½ | $1.672 \times 10^{-27}$ | $1.578 \times 10^{-16}$ | -e | c |

[a]All fermions in the self-system, [b]Composite particle